\documentclass[showpacs,amsmath,amssymb,twocolumn,prl,superscriptaddress]{revtex4-1}
\usepackage{amssymb}
\usepackage[dvips]{graphicx}
\usepackage{enumerate}
\usepackage{epsfig}
\usepackage{subfigure}
\usepackage{xcolor}
\usepackage[T1]{fontenc}
\usepackage{fullpage}
\usepackage{amsthm,amsfonts,amssymb,amscd,mathrsfs,xspace,framed}
\usepackage{amsmath}
\usepackage{color}
\usepackage{setspace}
\usepackage{url}
\usepackage{wrapfig}
\usepackage{enumitem}
\begin{document}

\title{Photonic analogue of Mollow triplet with on-chip photon pair generation}

\author{Chaohan Cui}
\affiliation{James C. Wyant College of Optical Sciences, The University of Arizona, Tucson, Arizona 85721, USA}
\author{Liang Zhang}
\affiliation{James C. Wyant College of Optical Sciences, The University of Arizona, Tucson, Arizona 85721, USA}
\author{Linran Fan}
\email{lfan@optics.arizona.edu}
\affiliation{James C. Wyant College of Optical Sciences, The University of Arizona, Tucson, Arizona 85721, USA}

\maketitle

\textbf{
The spontaneous emission of photons from a ladder of dressed state pairs leads to the prominent Mollow triplet in atomic systems \cite{mollow1969power, schuda1974observation, aspect1980time}. Here we propose and demonstrate the photonic analogue of Mollow triplet in the quantum regime. Photonic entanglement is generated with spontaneous nonlinear processes in dressed photonic modes, which are introduced through coherent multimode coupling. The photonic Mollow triplet is observed through the measurement of quantum temporal correlation between entangled photons. We further demonstrate the flexibility of the photonic system to realize different configurations of dressed states, leading to the controlled modification of Mollow triplet. Our work would enable the investigation of complex atomic processes and the realization of unique quantum functionalities based on photonic systems.
}

The modification of internal energy levels can lead to the dramatic change of the optical response in atoms and molecules. As an example, one central peak with two symmetric sidebands can be observed in the fluorescence spectrum of atomic systems driven by a resonant optical field due to the energy splitting between dressed states, which is known as Mollow triplet \cite{mollow1969power}. Along with Rabi oscillation, it is regarded as an important confirmation of light-atom interaction model in quantum optics  \cite{schuda1974observation,aspect1980time}. Recently, the study of Mollow triplet has attracted a renewed interest due to its potential applications in quantum information science \cite{baur2009measurement,flagg2009resonantly,ulhaq2012cascaded,pigeau2015observation,fischer2016self,lagoudakis2017observation,joas2017quantum,meinel2020heterodyne}. Highlighted by the capability of tailoring quantum correlations, Mollow triplet shows great promise to engineering high-quality quantum emitters \cite{flagg2009resonantly,ulhaq2012cascaded}. Novel quantum sensing schemes have also been demonstrated based on Mollow triplet \cite{joas2017quantum,meinel2020heterodyne}. In addition, Mollow triplet has been used to study the coherent dynamics of solid-state qubit systems such as quantum dots \cite{flagg2009resonantly,fischer2016self,lagoudakis2017observation}, superconducting Josephson junctions \cite{baur2009measurement}, and diamond color centers \cite{pigeau2015observation}.

The analogy between photonic resonances and atomic energy levels has enabled the investigation of complex photonic processes with established concepts of atomic physics. Prominent examples include photonic crystals \cite{joannopoulos1997photonic}, topological photonic systems \cite{ozawa2019topological}, and parity–time symmetric systems \cite{ozdemir2019parity}. It has led to the development of important photonic technologies including optical memories \cite{safavi2011electromagnetically, weis2010optomechanically, fan2015cascaded}, topologically protected laser \cite{feng2014single,st2017lasing}, structured light \cite{bahari2021photonic}, etc. Atomic processes such as quantum Hall effect \cite{wang2009observation,bahari2021photonic}, electromagnetically induced transparency \cite{safavi2011electromagnetically, weis2010optomechanically, fan2015cascaded}, and Ramsey interference \cite{zhang2019electronically}, have all been realized with various photonic platforms. While quantum interpretation is required to model atomic systems, it is sufficient to describe the corresponding photonic processes in the classical picture. Therefore, experimental demonstrations are also conducted with coherent optical fields \cite{safavi2011electromagnetically,weis2010optomechanically,fan2015cascaded,feng2014single,st2017lasing,bahari2021photonic,wang2009observation,zhang2019electronically}. The further expansion of the analogy between photonic and atomic systems into quantum regime will be highly valuable for the future development of photonic quantum information.

\begin{figure*}[htbp]
\centering
\includegraphics[width=5.25in]{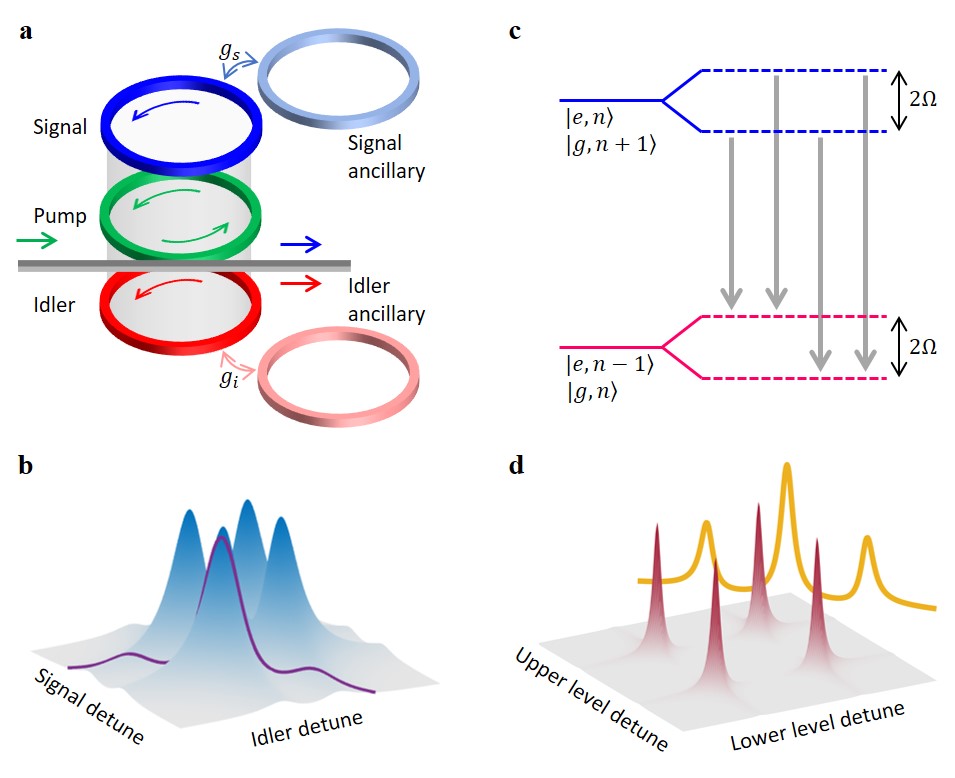}
\caption{\textbf{Quantum photonic Mollow triplet.} \textbf{a.}  Schematic to realize Mollow triplet with quantum photonic systems. In a photonic ring cavity with coherent drive (green), one signal (blue) and one idler (red) photons are created from spontaneous nonlinear processes. The signal and idler modes are further coupled to two ancillary modes respectively (light blue and pink). \textbf{b.} Joint spectral density of signal-idler generation. The purple line shows the case with a continuous-wave pump and non-zero dispersion. We assume equal coupling strengths for signal and idler modes $g_s=g_i$. \textbf{c.} The energy-level diagram for the atomic system under resonant optical drive. The degeneracy is lifted between $|e,n\rangle$ and $|g,n+1\rangle$ with $|e\rangle$ ($|g\rangle$) the atomic excited (ground) state and $|n\rangle$ the photon number state. Rabi frequency is $\Omega$. Four transition paths are labeled with grey arrows. \textbf{d.} Probability density for atomic transitions. The integral along diagonal direction (orange) is proportional to the fluorescence spectrum.}
\label{fig:Fig1} 
\end{figure*}

\begin{figure*}[htbp]
\centering
\includegraphics[width=6in]{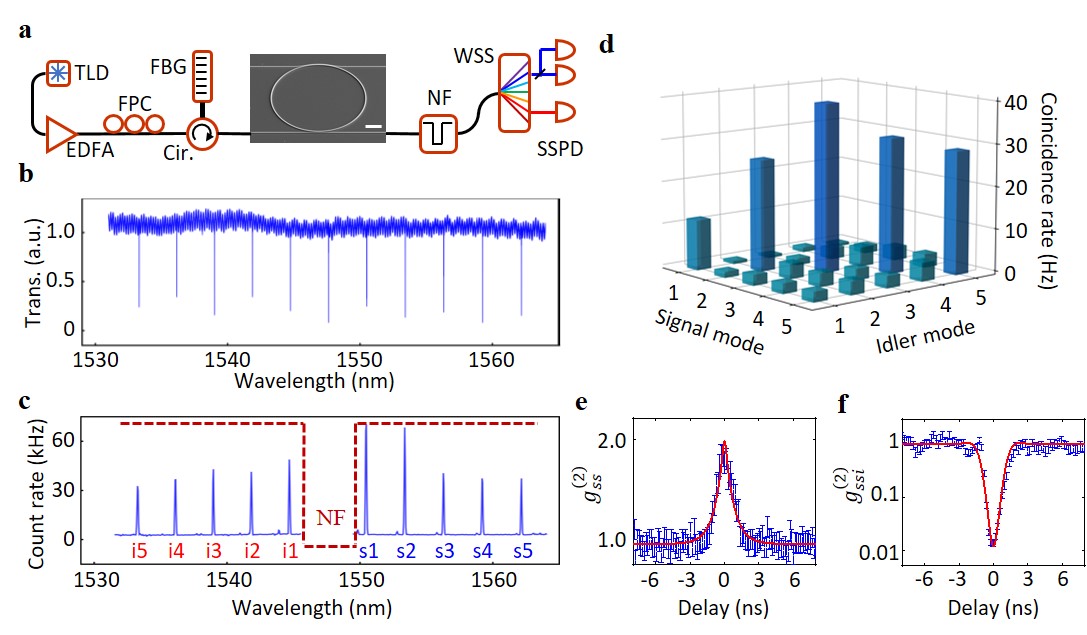}
\caption{\textbf{Entangled photon pair generation.} \textbf{a.} The measurement setup. TLD, tunable laser diode; EDFA, erbium-doped fiber amplifier; FPC, fiber polarization controller; Cir., circulator; FBG, fiber Bragg grating; NF, notch filter; WSS, wavelength selective switch; SSPD, superconducting single-photon detector. The scale bar in the scanning electron microscope image of the device is 20 $\mu$m. \textbf{b.} Transmission spectrum of the photonic ring cavity. \textbf{c.} Single-photon spectrum. The signal (idler) modes are labeled as s1-s5 (i1-i5). The notch filter (NF) regime is labeled in red. \textbf{d.} Coincidence rates measured at all the signal/idler combinations. Strong coincidence is only visible between symmetric signal-idler modes. \textbf{e.} Self-correlation function of signal s4 mode with no heralding. The data (blue) is fitted with double-exponential decay (red). The fitted peak value is $g_{ss}^{(2)} (0)=2.01\pm0.07$, which agrees the theoretical value  $g_{ss}^{(2)} (0)=2$ for a single-mode thermal state. \textbf{f.} Self-correlation measurement of signal s4 mode with idler i4 mode as heralding. The fitted dip value is $g_{ssi}^{(2)} (0)=0.009\pm0.007\ll0.5$, showing single photon generation.}
\label{fig:Fig2} 
\end{figure*}

Here, we demonstrate Mollow triplet with all-photonic systems in the quantum regime. With entanglement generation, the proposed photonic process cannot be interpreted in the classical picture, nor realized with coherent optical fields. The schematic of the photonic system is shown in Fig.\;\ref{fig:Fig1}a. We start with a photonic ring cavity evanescently coupled to a bus waveguide. A coherent drive is launched into the cavity to initiate the spontaneous nonlinear optical process (such as parametric down-conversion and spontaneous four-wave mixing). Non-degenerate correlated photon pairs can be  generated in symmetric signal and idler whispering gallery modes satisfying energy conservation. While a series of optical whispering gallery modes are supported in the photonic cavity, we focus on one signal-idler pair ($\hat{a}_s$ and $\hat{a}_i$). At the same time, the signal and idler modes are coherently coupled to the ancillary signal and idler modes ($\hat{b}_s$ and $\hat{b}_i$) respectively. The coherent coupling can be realized by additional pumps in nonlinear optical processes \cite{guo2018all}, microwave drives in electro-optic devices \cite{zhang2019electronically}, evanescently coupled photonic resonators \cite{peng2014and}, etc. We use the following Hamiltonian to describe the coupled multimode photonic system:
\begin{equation}
\begin{aligned}
    H=&\sum_{k\in{(s,i})}\hbar \omega_k (\hat{a}_k^\dagger \hat{a}_k + \hat{b}_k^\dagger \hat{b}_k) + 
    \hbar G (\hat{a}_s^\dagger \hat{a}_i^\dagger + \hat{a}_s \hat{a}_i) + \\
    &\hbar g_s (\hat{a}_s^\dagger \hat{b}_s + \hat{a}_s \hat{b}_s^\dagger) + 
    \hbar g_i (\hat{a}_i^\dagger \hat{b}_i + \hat{a}_i \hat{b}_i^\dagger).
\end{aligned}
\label{eq:sp}
\end{equation}
Here $G$ is the pump-enhanced coupling rate for the parametric nonlinear process, $g_s$ ($g_i$) is the coherent coupling strength between the signal (idler) mode and the corresponding ancillary mode, and $\omega_k$ is the angular frequency for mode $k$ with $k\in{(s,i})$. We also assume the signal (idler) mode and the corresponding ancillary mode have the same resonance frequency $\omega_s$ ($\omega_i$).

The coherent coupling between the signal (idler) mode and the corresponding ancillary mode leads to the re-normalization of the Hamiltonian, thus the creation of the dressed signal (idler) modes $\hat{a}_s\pm\hat{b}_s$ ($\hat{a}_i\pm\hat{b}_i$). Photon pair generation can occur with four transition paths with different combinations of the dressed signal and idler modes, shown as the four peaks in the joint spectral density (Fig.\;\ref{fig:Fig1}b). If a broadband pump is used, all four transition paths can be efficiently excited. Therefore, the biphoton spectrum is proportional to the integral of the joint spectral density along the diagonal direction. With equal coupling strengths $g_s=g_i$, two transition paths overlap in frequency. Therefore, one central peak with two sidebands shows up in the spectrum, which is the same as atomic Mollow triplet. If a continuous-wave pump is used, the energy conservation requires that the sum of signal and idler frequencies is constant. Therefore, the transition can only happen along the anti-diagonal direction. With proper dispersion, triplet structure can also be observed (purple line in Fig.\;\ref{fig:Fig1}b, and more cases in Supplementary Information).

In comparison, we also present the energy diagram of atomic systems resonantly driven with an optical field (Fig.\;\ref{fig:Fig1}c). Dressed states are generated due to the coherent coupling between degenerate states, leading to four transition paths for spontaneous photon emission. The fluorescence spectrum is proportional to the integral of transition probability density along the diagonal direction. Therefore, Mollow triplet with a central peak and two sidebands can be observed (Fig.\;\ref{fig:Fig1}d).

In our experiment, we use integrated ring resonators fabricated from single-crystal aluminum nitride (AlN) on sapphire to demonstrate photonic Mollow triplet (Fig.\;\ref{fig:Fig2}a and Methods). Entangled photon pairs are generated with spontaneous four-wave mixing by launching a continuous-wave pump into the resonance near 1547.60\;nm (Fig.\;\ref{fig:Fig2}b). With broadband phase matching, photons are generated in multiple modes spectrally symmetric to the pump resonance (Fig.\;\ref{fig:Fig2}c). With energy conservation, strong frequency correlation is only observed between symmetric signal and idler modes (Fig.\;\ref{fig:Fig2}d). We further isolate one signal mode, and measure the corresponding self-correlation function (Fig.\;\ref{fig:Fig2}e). As the idler photon is discarded, the signal photon is in the single-mode thermal state showing bunching statistics. We obtain  $g_{ss}^{(2)} (0)=2.01\pm0.07$, which is in agreement with the expected value  $g_{ss}^{(2)} (0)=2$ for a single-mode thermal state \cite{christ2011probing}. With the strong frequency correlation, we can further use the corresponding idler photon to herald the presence of the signal photon. As shown in Fig.\;\ref{fig:Fig2}f, the heralded self-correlation $g_{ssi}^{(2)} (0)$ shows non-classical anti-bunching behavior, with minimum value close to zero. This confirms that single photons are generated from the photonic ring cavity.

\begin{figure}[tb]
\centering
\includegraphics[width=3.02in]{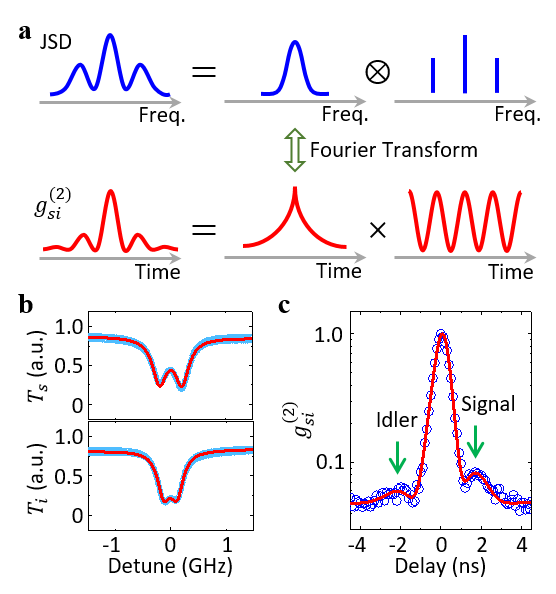}
\caption{\textbf{Observation of Mollow triplet with quantum temporal correlations.} \textbf{a.} Schematic showing that the signal-idler temporal correlation is the Fourier transform of the joint spectrum density. Mollow triplet is equivalent to the convolution between one Lorentzian shape and three equal-distance delta functions in frequency (blue). In the rotating frame of the central carrier frequency, the Fourier transforms of the Lorentzian shape and three equal-distance delta functions are double-exponential decay and sinusoidal functions respectively. Therefore, the temporal correlation is the product of double-exponential decay and sine wave (red). \textbf{b.} Linear transmission of signal s1 and idler i1 modes (cyan). Coupling strengths  $g_s=219$ MHz and $g_i=131$ MHz are obtained through fitting with double Lorentzian shape (red). \textbf{c.} Temporal correlation between signal s1 and idler s1 modes (blue) shows two sidebands introduced by signal and idler coherent coupling with ancillary modes respectively. The theoretical temporal correlation function is shown in red. This proves Mollow triplet structure in the joint spectral density.}
\label{fig:Fig3} 
\end{figure}

The photonic ring cavity supports both clock-wise (CW) and counter clock-wise (CCW) whispering gallery modes, corresponding to the original signal-idler and ancillary modes respectively. The coherent coupling between CW and CCW modes is introduced by the coherent back-scattering in the ring cavity \cite{li2016backscattering}. Therefore, entangled photon pairs are generated in the dressed states consisting of CW and CCW modes. Temporal correlation function between signal and idler photons $g_{si}^{(2)}$ is measured to verify photonic Mollow triplet. As shown in Fig.\;\ref{fig:Fig3}a, Mollow triplet in joint spectral density can be considered as the convolution between a Lorentzian function with three equal-distance delta functions. The temporal correlation $g_{si}^{(2)}$ is proportional to the Fourier transform of joint spectral density, thus the product of double-exponential decay and sinusoidal function (Supplementary Information). As a result, sidebands can also be observed in the temporal correlation function (Fig.\;\ref{fig:Fig3}a). From the linear transmission, we can clearly observe the mode splitting for both signal and idler modes, indicating that photonic dressed states are formed with the coherent superposition of CW and CCW modes (Fig.\;\ref{fig:Fig3}b). From the mode splitting amplitude, we can estimate the coupling strengths for the signal ($g_s=219$ MHz) and idler ($g_i=131$ MHz) modes respectively. The temporal correlation function $g_{si}^{(2)}$ is shown in Fig.\;\ref{fig:Fig3}c. In addition to the central peak, two sidebands are clearly observed, proving Mollow triplet structure in the joint spectral density. We further notice that the two sidebands are asymmetric in term of both strength and the frequency separation from the central peak. This is caused by the difference in the signal and idler coupling strengths.

\begin{figure}[tb]
\centering
\includegraphics[width=2.97in]{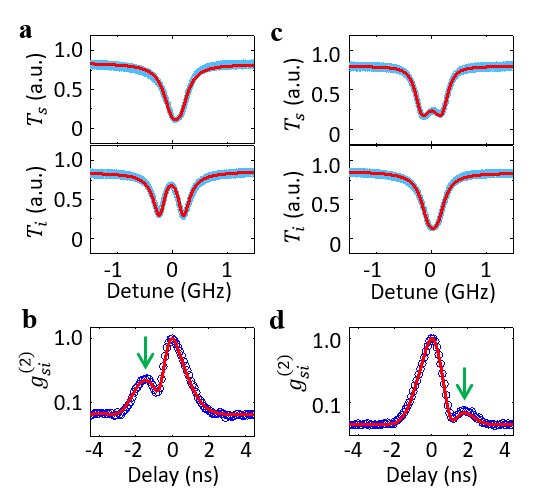}
\caption{\textbf{Mollow phenomena with single sideband.} \textbf{a.} Transmission spectrum of signal s2 and idler i2 modes (cyan). Only the idler mode shows coherent coupling with $g_i=247$ MHz. \textbf{b.} Measured temporal correlation between signal s2 and idler i2 modes (blue). Only the left sideband is observed. \textbf{c.} Transmission spectrum of signal s3 and idler i3 modes (cyan). Only the signal mode shows coherent coupling with $g_s=175$ MHz. \textbf{d.} Measured temporal correlation between signal s3 and idler i3 modes (blue). Only the right sideband is observed. Coupling strengths are obtained through fitting linear transmission with double Lorentzian shape (red) in \textbf{a} and \textbf{c}. The theoretical temporal correlation functions are shown in red in \textbf{b} and \textbf{d}.}
\label{fig:Fig4} 
\end{figure}

As each photonic mode can be engineered independently, photonic systems can provide more flexibility in exploring corresponding physics processes than atomic systems. For example, Mollow triplet with atomic systems is usually symmetric, as the energy splitting is the same for all energy levels \cite{mollow1969power, schuda1974observation, aspect1980time}. With photonic systems, we can turn off the coherent coupling for the signal mode, and realize strong coupling for the idler mode (Fig.\;\ref{fig:Fig4}a). In this case, only the left sideband is observed instead of two symmetric ones (Fig.\;\ref{fig:Fig4}b). Similarly, we can also realize strong and zero coupling strengths for the signal and idler modes respectively (Fig.\;\ref{fig:Fig4}c). By switching the coupling strengths between signal and idler modes, we effectively conduct a reflection operation in the time domain. Therefore, only the right sideband is observed in the temporal correlation function (Fig.\;\ref{fig:Fig4}d).

In conclusion, we report the demonstration of Mollow triplet with all-photonic systems using entangled photons. This work expands the analogy between atomic and photonic systems into the quantum regime. It will enable the development of novel quantum photonic functions such as quantum spectro-temporal shaping \cite{ansari2018tailoring,karpinski2017bandwidth,fan2016integrated} and high-dimensional quantum information processing \cite{kues2019quantum}. Advanced coupling schemes such as dynamic coupling control with nonlinear optics \cite{guo2018all} and microwave drives \cite{zhang2019electronically} can further unlock the capability to study complex physics processes with photons.

\vbox{}
\textbf{Methods}

The device is fabricated from single-crystal aluminum nitride (AlN) grown on sapphire substrate with metal organic chemical vapor deposition (MOCVD). Photonic circuits are patterned with electron-beam lithography (EBL) using hydrogen silsesquioxane (HSQ) resist, which are subsequently transferred to the AlN layer by chlorine-based reactive-ion etching (RIE). Then SiO$_2$ layer is deposited as the top cladding. Nominal thickness of AlN layer is 1 $\mu$m. The ring cavity has waveguide width 1.2 $\mu$m and radius 60 $\mu$m. Therefore, the ring cavity shows strong anomalous dispersion $D_\lambda\approx80\;$ps/(km$\cdot$nm) near 1550 nm wavelength.

The light from a tunable laser diode is amplified by an erbium-doped fiber amplifier (EDFA), then amplified spontaneous emission (ASE) noise is removed by bandpass filter consisted of fiber Bragg gratings (FBG) and optical circulators (Cir.). The light is launched into and collected from the device with a pair of lensed fibers. A series of notch filters is used to remove the pump. Then the collected light is sent into wavelength-selective switch to isolate different signal and idler photons, which are detected by individual superconducting single photon detectors. The photon arrival time is registered with a time tagger. 

\vbox{}
\textbf{Acknowledgments}

This work was supported by Office of Naval Research (N00014-19-1-2190), National Science Foundation (ECCS-1842559, CCF-1907918), and U.S. Department of Energy UT-Battelle/Oak Ridge National Laboratory (4000178321).

\bibliography{Ref}

\begin{thebibliography}{30}
\expandafter\ifx\csname natexlab\endcsname\relax\def\natexlab#1{#1}\fi
\expandafter\ifx\csname bibnamefont\endcsname\relax
  \def\bibnamefont#1{#1}\fi
\expandafter\ifx\csname bibfnamefont\endcsname\relax
  \def\bibfnamefont#1{#1}\fi
\expandafter\ifx\csname citenamefont\endcsname\relax
  \def\citenamefont#1{#1}\fi
\expandafter\ifx\csname url\endcsname\relax
  \def\url#1{\texttt{#1}}\fi
\expandafter\ifx\csname urlprefix\endcsname\relax\def\urlprefix{URL }\fi
\providecommand{\bibinfo}[2]{#2}
\providecommand{\eprint}[2][]{\url{#2}}

\bibitem[{\citenamefont{Mollow}(1969)}]{mollow1969power}
\bibinfo{author}{\bibfnamefont{B.}~\bibnamefont{Mollow}},
  \bibinfo{journal}{Physical Review} \textbf{\bibinfo{volume}{188}}
  (\bibinfo{year}{1969}).

\bibitem[{\citenamefont{Schuda et~al.}(1974)\citenamefont{Schuda, Stroud~Jr,
  and Hercher}}]{schuda1974observation}
\bibinfo{author}{\bibfnamefont{F.}~\bibnamefont{Schuda}},
  \bibinfo{author}{\bibfnamefont{C.}~\bibnamefont{Stroud~Jr}},
  \bibnamefont{and} \bibinfo{author}{\bibfnamefont{M.}~\bibnamefont{Hercher}},
  \bibinfo{journal}{Journal of Physics B: Atomic and Molecular Physics}
  \textbf{\bibinfo{volume}{7}}, \bibinfo{pages}{L198} (\bibinfo{year}{1974}).

\bibitem[{\citenamefont{Aspect et~al.}(1980)\citenamefont{Aspect, Roger,
  Reynaud, Dalibard, and Cohen-Tannoudji}}]{aspect1980time}
\bibinfo{author}{\bibfnamefont{A.}~\bibnamefont{Aspect}},
  \bibinfo{author}{\bibfnamefont{G.}~\bibnamefont{Roger}},
  \bibinfo{author}{\bibfnamefont{S.}~\bibnamefont{Reynaud}},
  \bibinfo{author}{\bibfnamefont{J.}~\bibnamefont{Dalibard}}, \bibnamefont{and}
  \bibinfo{author}{\bibfnamefont{C.}~\bibnamefont{Cohen-Tannoudji}},
  \bibinfo{journal}{Physical Review Letters} \textbf{\bibinfo{volume}{45}},
  \bibinfo{pages}{617} (\bibinfo{year}{1980}).

\bibitem[{\citenamefont{Baur et~al.}(2009)\citenamefont{Baur, Filipp,
  Bianchetti, Fink, G{\"o}ppl, Steffen, Leek, Blais, and
  Wallraff}}]{baur2009measurement}
\bibinfo{author}{\bibfnamefont{M.}~\bibnamefont{Baur}},
  \bibinfo{author}{\bibfnamefont{S.}~\bibnamefont{Filipp}},
  \bibinfo{author}{\bibfnamefont{R.}~\bibnamefont{Bianchetti}},
  \bibinfo{author}{\bibfnamefont{J.}~\bibnamefont{Fink}},
  \bibinfo{author}{\bibfnamefont{M.}~\bibnamefont{G{\"o}ppl}},
  \bibinfo{author}{\bibfnamefont{L.}~\bibnamefont{Steffen}},
  \bibinfo{author}{\bibfnamefont{P.~J.} \bibnamefont{Leek}},
  \bibinfo{author}{\bibfnamefont{A.}~\bibnamefont{Blais}}, \bibnamefont{and}
  \bibinfo{author}{\bibfnamefont{A.}~\bibnamefont{Wallraff}},
  \bibinfo{journal}{Physical Review Letters} \textbf{\bibinfo{volume}{102}},
  \bibinfo{pages}{243602} (\bibinfo{year}{2009}).

\bibitem[{\citenamefont{Flagg et~al.}(2009)\citenamefont{Flagg, Muller,
  Robertson, Founta, Deppe, Xiao, Ma, Salamo, and Shih}}]{flagg2009resonantly}
\bibinfo{author}{\bibfnamefont{E.~B.} \bibnamefont{Flagg}},
  \bibinfo{author}{\bibfnamefont{A.}~\bibnamefont{Muller}},
  \bibinfo{author}{\bibfnamefont{J.}~\bibnamefont{Robertson}},
  \bibinfo{author}{\bibfnamefont{S.}~\bibnamefont{Founta}},
  \bibinfo{author}{\bibfnamefont{D.}~\bibnamefont{Deppe}},
  \bibinfo{author}{\bibfnamefont{M.}~\bibnamefont{Xiao}},
  \bibinfo{author}{\bibfnamefont{W.}~\bibnamefont{Ma}},
  \bibinfo{author}{\bibfnamefont{G.}~\bibnamefont{Salamo}}, \bibnamefont{and}
  \bibinfo{author}{\bibfnamefont{C.-K.} \bibnamefont{Shih}},
  \bibinfo{journal}{Nature Physics} \textbf{\bibinfo{volume}{5}},
  \bibinfo{pages}{203} (\bibinfo{year}{2009}).

\bibitem[{\citenamefont{Ulhaq et~al.}(2012)\citenamefont{Ulhaq, Weiler, Ulrich,
  Ro{\ss}bach, Jetter, and Michler}}]{ulhaq2012cascaded}
\bibinfo{author}{\bibfnamefont{A.}~\bibnamefont{Ulhaq}},
  \bibinfo{author}{\bibfnamefont{S.}~\bibnamefont{Weiler}},
  \bibinfo{author}{\bibfnamefont{S.~M.} \bibnamefont{Ulrich}},
  \bibinfo{author}{\bibfnamefont{R.}~\bibnamefont{Ro{\ss}bach}},
  \bibinfo{author}{\bibfnamefont{M.}~\bibnamefont{Jetter}}, \bibnamefont{and}
  \bibinfo{author}{\bibfnamefont{P.}~\bibnamefont{Michler}},
  \bibinfo{journal}{Nature Photonics} \textbf{\bibinfo{volume}{6}},
  \bibinfo{pages}{238} (\bibinfo{year}{2012}).

\bibitem[{\citenamefont{Pigeau et~al.}(2015)\citenamefont{Pigeau, Rohr,
  De~L{\'e}pinay, Gloppe, Jacques, and Arcizet}}]{pigeau2015observation}
\bibinfo{author}{\bibfnamefont{B.}~\bibnamefont{Pigeau}},
  \bibinfo{author}{\bibfnamefont{S.}~\bibnamefont{Rohr}},
  \bibinfo{author}{\bibfnamefont{L.~M.} \bibnamefont{De~L{\'e}pinay}},
  \bibinfo{author}{\bibfnamefont{A.}~\bibnamefont{Gloppe}},
  \bibinfo{author}{\bibfnamefont{V.}~\bibnamefont{Jacques}}, \bibnamefont{and}
  \bibinfo{author}{\bibfnamefont{O.}~\bibnamefont{Arcizet}},
  \bibinfo{journal}{Nature Communications} \textbf{\bibinfo{volume}{6}},
  \bibinfo{pages}{1} (\bibinfo{year}{2015}).

\bibitem[{\citenamefont{Fischer et~al.}(2016)\citenamefont{Fischer, M{\"u}ller,
  Rundquist, Sarmiento, Piggott, Kelaita, Dory, Lagoudakis, and
  Vu{\v{c}}kovi{\'c}}}]{fischer2016self}
\bibinfo{author}{\bibfnamefont{K.~A.} \bibnamefont{Fischer}},
  \bibinfo{author}{\bibfnamefont{K.}~\bibnamefont{M{\"u}ller}},
  \bibinfo{author}{\bibfnamefont{A.}~\bibnamefont{Rundquist}},
  \bibinfo{author}{\bibfnamefont{T.}~\bibnamefont{Sarmiento}},
  \bibinfo{author}{\bibfnamefont{A.~Y.} \bibnamefont{Piggott}},
  \bibinfo{author}{\bibfnamefont{Y.}~\bibnamefont{Kelaita}},
  \bibinfo{author}{\bibfnamefont{C.}~\bibnamefont{Dory}},
  \bibinfo{author}{\bibfnamefont{K.~G.} \bibnamefont{Lagoudakis}},
  \bibnamefont{and}
  \bibinfo{author}{\bibfnamefont{J.}~\bibnamefont{Vu{\v{c}}kovi{\'c}}},
  \bibinfo{journal}{Nature Photonics} \textbf{\bibinfo{volume}{10}},
  \bibinfo{pages}{163} (\bibinfo{year}{2016}).

\bibitem[{\citenamefont{Lagoudakis et~al.}(2017)\citenamefont{Lagoudakis,
  Fischer, Sarmiento, McMahon, Radulaski, Zhang, Kelaita, Dory, M{\"u}ller, and
  Vu{\v{c}}kovi{\'c}}}]{lagoudakis2017observation}
\bibinfo{author}{\bibfnamefont{K.}~\bibnamefont{Lagoudakis}},
  \bibinfo{author}{\bibfnamefont{K.}~\bibnamefont{Fischer}},
  \bibinfo{author}{\bibfnamefont{T.}~\bibnamefont{Sarmiento}},
  \bibinfo{author}{\bibfnamefont{P.}~\bibnamefont{McMahon}},
  \bibinfo{author}{\bibfnamefont{M.}~\bibnamefont{Radulaski}},
  \bibinfo{author}{\bibfnamefont{J.}~\bibnamefont{Zhang}},
  \bibinfo{author}{\bibfnamefont{Y.}~\bibnamefont{Kelaita}},
  \bibinfo{author}{\bibfnamefont{C.}~\bibnamefont{Dory}},
  \bibinfo{author}{\bibfnamefont{K.}~\bibnamefont{M{\"u}ller}},
  \bibnamefont{and}
  \bibinfo{author}{\bibfnamefont{J.}~\bibnamefont{Vu{\v{c}}kovi{\'c}}},
  \bibinfo{journal}{Physical Review Letters} \textbf{\bibinfo{volume}{118}},
  \bibinfo{pages}{013602} (\bibinfo{year}{2017}).

\bibitem[{\citenamefont{Joas et~al.}(2017)\citenamefont{Joas, Waeber,
  Braunbeck, and Reinhard}}]{joas2017quantum}
\bibinfo{author}{\bibfnamefont{T.}~\bibnamefont{Joas}},
  \bibinfo{author}{\bibfnamefont{A.~M.} \bibnamefont{Waeber}},
  \bibinfo{author}{\bibfnamefont{G.}~\bibnamefont{Braunbeck}},
  \bibnamefont{and} \bibinfo{author}{\bibfnamefont{F.}~\bibnamefont{Reinhard}},
  \bibinfo{journal}{Nature Communications} \textbf{\bibinfo{volume}{8}},
  \bibinfo{pages}{1} (\bibinfo{year}{2017}).

\bibitem[{\citenamefont{Meinel et~al.}(2020)\citenamefont{Meinel, Vorobyov,
  Yavkin, Dasari, Sumiya, Onoda, Isoya, and Wrachtrup}}]{meinel2020heterodyne}
\bibinfo{author}{\bibfnamefont{J.}~\bibnamefont{Meinel}},
  \bibinfo{author}{\bibfnamefont{V.}~\bibnamefont{Vorobyov}},
  \bibinfo{author}{\bibfnamefont{B.}~\bibnamefont{Yavkin}},
  \bibinfo{author}{\bibfnamefont{D.}~\bibnamefont{Dasari}},
  \bibinfo{author}{\bibfnamefont{H.}~\bibnamefont{Sumiya}},
  \bibinfo{author}{\bibfnamefont{S.}~\bibnamefont{Onoda}},
  \bibinfo{author}{\bibfnamefont{J.}~\bibnamefont{Isoya}}, \bibnamefont{and}
  \bibinfo{author}{\bibfnamefont{J.}~\bibnamefont{Wrachtrup}},
  \bibinfo{journal}{arXiv preprint arXiv:2008.10068}  (\bibinfo{year}{2020}).

\bibitem[{\citenamefont{Joannopoulos et~al.}(1997)\citenamefont{Joannopoulos,
  Villeneuve, and Fan}}]{joannopoulos1997photonic}
\bibinfo{author}{\bibfnamefont{J.~D.} \bibnamefont{Joannopoulos}},
  \bibinfo{author}{\bibfnamefont{P.~R.} \bibnamefont{Villeneuve}},
  \bibnamefont{and} \bibinfo{author}{\bibfnamefont{S.}~\bibnamefont{Fan}},
  \bibinfo{journal}{Nature} \textbf{\bibinfo{volume}{386}},
  \bibinfo{pages}{143} (\bibinfo{year}{1997}).

\bibitem[{\citenamefont{Ozawa et~al.}(2019)\citenamefont{Ozawa, Price, Amo,
  Goldman, Hafezi, Lu, Rechtsman, Schuster, Simon, Zilberberg
  et~al.}}]{ozawa2019topological}
\bibinfo{author}{\bibfnamefont{T.}~\bibnamefont{Ozawa}},
  \bibinfo{author}{\bibfnamefont{H.~M.} \bibnamefont{Price}},
  \bibinfo{author}{\bibfnamefont{A.}~\bibnamefont{Amo}},
  \bibinfo{author}{\bibfnamefont{N.}~\bibnamefont{Goldman}},
  \bibinfo{author}{\bibfnamefont{M.}~\bibnamefont{Hafezi}},
  \bibinfo{author}{\bibfnamefont{L.}~\bibnamefont{Lu}},
  \bibinfo{author}{\bibfnamefont{M.~C.} \bibnamefont{Rechtsman}},
  \bibinfo{author}{\bibfnamefont{D.}~\bibnamefont{Schuster}},
  \bibinfo{author}{\bibfnamefont{J.}~\bibnamefont{Simon}},
  \bibinfo{author}{\bibfnamefont{O.}~\bibnamefont{Zilberberg}},
  \bibnamefont{et~al.}, \bibinfo{journal}{Reviews of Modern Physics}
  \textbf{\bibinfo{volume}{91}}, \bibinfo{pages}{015006}
  (\bibinfo{year}{2019}).

\bibitem[{\citenamefont{{\"O}zdemir et~al.}(2019)\citenamefont{{\"O}zdemir,
  Rotter, Nori, and Yang}}]{ozdemir2019parity}
\bibinfo{author}{\bibfnamefont{{\c{S}}.~K.} \bibnamefont{{\"O}zdemir}},
  \bibinfo{author}{\bibfnamefont{S.}~\bibnamefont{Rotter}},
  \bibinfo{author}{\bibfnamefont{F.}~\bibnamefont{Nori}}, \bibnamefont{and}
  \bibinfo{author}{\bibfnamefont{L.}~\bibnamefont{Yang}},
  \bibinfo{journal}{Nature Materials} \textbf{\bibinfo{volume}{18}},
  \bibinfo{pages}{783} (\bibinfo{year}{2019}).

\bibitem[{\citenamefont{Safavi-Naeini et~al.}(2011)\citenamefont{Safavi-Naeini,
  Alegre, Chan, Eichenfield, Winger, Lin, Hill, Chang, and
  Painter}}]{safavi2011electromagnetically}
\bibinfo{author}{\bibfnamefont{A.~H.} \bibnamefont{Safavi-Naeini}},
  \bibinfo{author}{\bibfnamefont{T.~M.} \bibnamefont{Alegre}},
  \bibinfo{author}{\bibfnamefont{J.}~\bibnamefont{Chan}},
  \bibinfo{author}{\bibfnamefont{M.}~\bibnamefont{Eichenfield}},
  \bibinfo{author}{\bibfnamefont{M.}~\bibnamefont{Winger}},
  \bibinfo{author}{\bibfnamefont{Q.}~\bibnamefont{Lin}},
  \bibinfo{author}{\bibfnamefont{J.~T.} \bibnamefont{Hill}},
  \bibinfo{author}{\bibfnamefont{D.~E.} \bibnamefont{Chang}}, \bibnamefont{and}
  \bibinfo{author}{\bibfnamefont{O.}~\bibnamefont{Painter}},
  \bibinfo{journal}{Nature} \textbf{\bibinfo{volume}{472}}, \bibinfo{pages}{69}
  (\bibinfo{year}{2011}).

\bibitem[{\citenamefont{Weis et~al.}(2010)\citenamefont{Weis, Rivi{\`e}re,
  Del{\'e}glise, Gavartin, Arcizet, Schliesser, and
  Kippenberg}}]{weis2010optomechanically}
\bibinfo{author}{\bibfnamefont{S.}~\bibnamefont{Weis}},
  \bibinfo{author}{\bibfnamefont{R.}~\bibnamefont{Rivi{\`e}re}},
  \bibinfo{author}{\bibfnamefont{S.}~\bibnamefont{Del{\'e}glise}},
  \bibinfo{author}{\bibfnamefont{E.}~\bibnamefont{Gavartin}},
  \bibinfo{author}{\bibfnamefont{O.}~\bibnamefont{Arcizet}},
  \bibinfo{author}{\bibfnamefont{A.}~\bibnamefont{Schliesser}},
  \bibnamefont{and} \bibinfo{author}{\bibfnamefont{T.~J.}
  \bibnamefont{Kippenberg}}, \bibinfo{journal}{Science}
  \textbf{\bibinfo{volume}{330}}, \bibinfo{pages}{1520} (\bibinfo{year}{2010}).

\bibitem[{\citenamefont{Fan et~al.}(2015)\citenamefont{Fan, Fong, Poot, and
  Tang}}]{fan2015cascaded}
\bibinfo{author}{\bibfnamefont{L.}~\bibnamefont{Fan}},
  \bibinfo{author}{\bibfnamefont{K.~Y.} \bibnamefont{Fong}},
  \bibinfo{author}{\bibfnamefont{M.}~\bibnamefont{Poot}}, \bibnamefont{and}
  \bibinfo{author}{\bibfnamefont{H.~X.} \bibnamefont{Tang}},
  \bibinfo{journal}{Nature Communications} \textbf{\bibinfo{volume}{6}},
  \bibinfo{pages}{1} (\bibinfo{year}{2015}).

\bibitem[{\citenamefont{Feng et~al.}(2014)\citenamefont{Feng, Wong, Ma, Wang,
  and Zhang}}]{feng2014single}
\bibinfo{author}{\bibfnamefont{L.}~\bibnamefont{Feng}},
  \bibinfo{author}{\bibfnamefont{Z.~J.} \bibnamefont{Wong}},
  \bibinfo{author}{\bibfnamefont{R.-M.} \bibnamefont{Ma}},
  \bibinfo{author}{\bibfnamefont{Y.}~\bibnamefont{Wang}}, \bibnamefont{and}
  \bibinfo{author}{\bibfnamefont{X.}~\bibnamefont{Zhang}},
  \bibinfo{journal}{Science} \textbf{\bibinfo{volume}{346}},
  \bibinfo{pages}{972} (\bibinfo{year}{2014}).

\bibitem[{\citenamefont{St-Jean et~al.}(2017)\citenamefont{St-Jean, Goblot,
  Galopin, Lema{\^\i}tre, Ozawa, Le~Gratiet, Sagnes, Bloch, and
  Amo}}]{st2017lasing}
\bibinfo{author}{\bibfnamefont{P.}~\bibnamefont{St-Jean}},
  \bibinfo{author}{\bibfnamefont{V.}~\bibnamefont{Goblot}},
  \bibinfo{author}{\bibfnamefont{E.}~\bibnamefont{Galopin}},
  \bibinfo{author}{\bibfnamefont{A.}~\bibnamefont{Lema{\^\i}tre}},
  \bibinfo{author}{\bibfnamefont{T.}~\bibnamefont{Ozawa}},
  \bibinfo{author}{\bibfnamefont{L.}~\bibnamefont{Le~Gratiet}},
  \bibinfo{author}{\bibfnamefont{I.}~\bibnamefont{Sagnes}},
  \bibinfo{author}{\bibfnamefont{J.}~\bibnamefont{Bloch}}, \bibnamefont{and}
  \bibinfo{author}{\bibfnamefont{A.}~\bibnamefont{Amo}},
  \bibinfo{journal}{Nature Photonics} \textbf{\bibinfo{volume}{11}},
  \bibinfo{pages}{651} (\bibinfo{year}{2017}).

\bibitem[{\citenamefont{Bahari et~al.}(2021)\citenamefont{Bahari, Hsu, Pan,
  Preece, Ndao, El~Amili, Fainman, and Kant{\'e}}}]{bahari2021photonic}
\bibinfo{author}{\bibfnamefont{B.}~\bibnamefont{Bahari}},
  \bibinfo{author}{\bibfnamefont{L.}~\bibnamefont{Hsu}},
  \bibinfo{author}{\bibfnamefont{S.~H.} \bibnamefont{Pan}},
  \bibinfo{author}{\bibfnamefont{D.}~\bibnamefont{Preece}},
  \bibinfo{author}{\bibfnamefont{A.}~\bibnamefont{Ndao}},
  \bibinfo{author}{\bibfnamefont{A.}~\bibnamefont{El~Amili}},
  \bibinfo{author}{\bibfnamefont{Y.}~\bibnamefont{Fainman}}, \bibnamefont{and}
  \bibinfo{author}{\bibfnamefont{B.}~\bibnamefont{Kant{\'e}}},
  \bibinfo{journal}{Nature Physics} pp. \bibinfo{pages}{1--4}
  (\bibinfo{year}{2021}).

\bibitem[{\citenamefont{Wang et~al.}(2009)\citenamefont{Wang, Chong,
  Joannopoulos, and Solja{\v{c}}i{\'c}}}]{wang2009observation}
\bibinfo{author}{\bibfnamefont{Z.}~\bibnamefont{Wang}},
  \bibinfo{author}{\bibfnamefont{Y.}~\bibnamefont{Chong}},
  \bibinfo{author}{\bibfnamefont{J.~D.} \bibnamefont{Joannopoulos}},
  \bibnamefont{and}
  \bibinfo{author}{\bibfnamefont{M.}~\bibnamefont{Solja{\v{c}}i{\'c}}},
  \bibinfo{journal}{Nature} \textbf{\bibinfo{volume}{461}},
  \bibinfo{pages}{772} (\bibinfo{year}{2009}).

\bibitem[{\citenamefont{Zhang et~al.}(2019)\citenamefont{Zhang, Wang, Hu,
  Shams-Ansari, Ren, Fan, and Lon{\v{c}}ar}}]{zhang2019electronically}
\bibinfo{author}{\bibfnamefont{M.}~\bibnamefont{Zhang}},
  \bibinfo{author}{\bibfnamefont{C.}~\bibnamefont{Wang}},
  \bibinfo{author}{\bibfnamefont{Y.}~\bibnamefont{Hu}},
  \bibinfo{author}{\bibfnamefont{A.}~\bibnamefont{Shams-Ansari}},
  \bibinfo{author}{\bibfnamefont{T.}~\bibnamefont{Ren}},
  \bibinfo{author}{\bibfnamefont{S.}~\bibnamefont{Fan}}, \bibnamefont{and}
  \bibinfo{author}{\bibfnamefont{M.}~\bibnamefont{Lon{\v{c}}ar}},
  \bibinfo{journal}{Nature Photonics} \textbf{\bibinfo{volume}{13}},
  \bibinfo{pages}{36} (\bibinfo{year}{2019}).

\bibitem[{\citenamefont{Guo et~al.}(2018)\citenamefont{Guo, Zou, Jiang, and
  Tang}}]{guo2018all}
\bibinfo{author}{\bibfnamefont{X.}~\bibnamefont{Guo}},
  \bibinfo{author}{\bibfnamefont{C.-L.} \bibnamefont{Zou}},
  \bibinfo{author}{\bibfnamefont{L.}~\bibnamefont{Jiang}}, \bibnamefont{and}
  \bibinfo{author}{\bibfnamefont{H.~X.} \bibnamefont{Tang}},
  \bibinfo{journal}{Physical Review Letters} \textbf{\bibinfo{volume}{120}},
  \bibinfo{pages}{203902} (\bibinfo{year}{2018}).

\bibitem[{\citenamefont{Peng et~al.}(2014)\citenamefont{Peng, {\"O}zdemir,
  Chen, Nori, and Yang}}]{peng2014and}
\bibinfo{author}{\bibfnamefont{B.}~\bibnamefont{Peng}},
  \bibinfo{author}{\bibfnamefont{{\c{S}}.~K.} \bibnamefont{{\"O}zdemir}},
  \bibinfo{author}{\bibfnamefont{W.}~\bibnamefont{Chen}},
  \bibinfo{author}{\bibfnamefont{F.}~\bibnamefont{Nori}}, \bibnamefont{and}
  \bibinfo{author}{\bibfnamefont{L.}~\bibnamefont{Yang}},
  \bibinfo{journal}{Nature Communications} \textbf{\bibinfo{volume}{5}},
  \bibinfo{pages}{1} (\bibinfo{year}{2014}).

\bibitem[{\citenamefont{Christ et~al.}(2011)\citenamefont{Christ, Laiho,
  Eckstein, Cassemiro, and Silberhorn}}]{christ2011probing}
\bibinfo{author}{\bibfnamefont{A.}~\bibnamefont{Christ}},
  \bibinfo{author}{\bibfnamefont{K.}~\bibnamefont{Laiho}},
  \bibinfo{author}{\bibfnamefont{A.}~\bibnamefont{Eckstein}},
  \bibinfo{author}{\bibfnamefont{K.~N.} \bibnamefont{Cassemiro}},
  \bibnamefont{and}
  \bibinfo{author}{\bibfnamefont{C.}~\bibnamefont{Silberhorn}},
  \bibinfo{journal}{New Journal of Physics} \textbf{\bibinfo{volume}{13}},
  \bibinfo{pages}{033027} (\bibinfo{year}{2011}).

\bibitem[{\citenamefont{Li et~al.}(2016)\citenamefont{Li, Van~Vaerenbergh,
  De~Heyn, Bienstman, and Bogaerts}}]{li2016backscattering}
\bibinfo{author}{\bibfnamefont{A.}~\bibnamefont{Li}},
  \bibinfo{author}{\bibfnamefont{T.}~\bibnamefont{Van~Vaerenbergh}},
  \bibinfo{author}{\bibfnamefont{P.}~\bibnamefont{De~Heyn}},
  \bibinfo{author}{\bibfnamefont{P.}~\bibnamefont{Bienstman}},
  \bibnamefont{and} \bibinfo{author}{\bibfnamefont{W.}~\bibnamefont{Bogaerts}},
  \bibinfo{journal}{Laser \& Photonics Reviews} \textbf{\bibinfo{volume}{10}},
  \bibinfo{pages}{420} (\bibinfo{year}{2016}).

\bibitem[{\citenamefont{Ansari et~al.}(2018)\citenamefont{Ansari, Donohue,
  Brecht, and Silberhorn}}]{ansari2018tailoring}
\bibinfo{author}{\bibfnamefont{V.}~\bibnamefont{Ansari}},
  \bibinfo{author}{\bibfnamefont{J.~M.} \bibnamefont{Donohue}},
  \bibinfo{author}{\bibfnamefont{B.}~\bibnamefont{Brecht}}, \bibnamefont{and}
  \bibinfo{author}{\bibfnamefont{C.}~\bibnamefont{Silberhorn}},
  \bibinfo{journal}{Optica} \textbf{\bibinfo{volume}{5}}, \bibinfo{pages}{534}
  (\bibinfo{year}{2018}).

\bibitem[{\citenamefont{Karpi{\'n}ski et~al.}(2017)\citenamefont{Karpi{\'n}ski,
  Jachura, Wright, and Smith}}]{karpinski2017bandwidth}
\bibinfo{author}{\bibfnamefont{M.}~\bibnamefont{Karpi{\'n}ski}},
  \bibinfo{author}{\bibfnamefont{M.}~\bibnamefont{Jachura}},
  \bibinfo{author}{\bibfnamefont{L.~J.} \bibnamefont{Wright}},
  \bibnamefont{and} \bibinfo{author}{\bibfnamefont{B.~J.} \bibnamefont{Smith}},
  \bibinfo{journal}{Nature Photonics} \textbf{\bibinfo{volume}{11}},
  \bibinfo{pages}{53} (\bibinfo{year}{2017}).

\bibitem[{\citenamefont{Fan et~al.}(2016)\citenamefont{Fan, Zou, Poot, Cheng,
  Guo, Han, and Tang}}]{fan2016integrated}
\bibinfo{author}{\bibfnamefont{L.}~\bibnamefont{Fan}},
  \bibinfo{author}{\bibfnamefont{C.-L.} \bibnamefont{Zou}},
  \bibinfo{author}{\bibfnamefont{M.}~\bibnamefont{Poot}},
  \bibinfo{author}{\bibfnamefont{R.}~\bibnamefont{Cheng}},
  \bibinfo{author}{\bibfnamefont{X.}~\bibnamefont{Guo}},
  \bibinfo{author}{\bibfnamefont{X.}~\bibnamefont{Han}}, \bibnamefont{and}
  \bibinfo{author}{\bibfnamefont{H.~X.} \bibnamefont{Tang}},
  \bibinfo{journal}{Nature Photonics} \textbf{\bibinfo{volume}{10}},
  \bibinfo{pages}{766} (\bibinfo{year}{2016}).

\bibitem[{\citenamefont{Kues et~al.}(2019)\citenamefont{Kues, Reimer, Lukens,
  Munro, Weiner, Moss, and Morandotti}}]{kues2019quantum}
\bibinfo{author}{\bibfnamefont{M.}~\bibnamefont{Kues}},
  \bibinfo{author}{\bibfnamefont{C.}~\bibnamefont{Reimer}},
  \bibinfo{author}{\bibfnamefont{J.~M.} \bibnamefont{Lukens}},
  \bibinfo{author}{\bibfnamefont{W.~J.} \bibnamefont{Munro}},
  \bibinfo{author}{\bibfnamefont{A.~M.} \bibnamefont{Weiner}},
  \bibinfo{author}{\bibfnamefont{D.~J.} \bibnamefont{Moss}}, \bibnamefont{and}
  \bibinfo{author}{\bibfnamefont{R.}~\bibnamefont{Morandotti}},
  \bibinfo{journal}{Nature Photonics} \textbf{\bibinfo{volume}{13}},
  \bibinfo{pages}{170} (\bibinfo{year}{2019}).

\end{thebibliography}

\end{document}